\documentclass[aps,twocolumn,showpacs,superscriptaddress,groupedaddress,nofootinbib]{revtex4-1}
\usepackage{graphicx}
\usepackage{bm}       
\usepackage{amssymb}
\usepackage[toc,page]{appendix}
\usepackage{braket}
\newcommand{\del}{\partial}
\newcommand{\f}{\frac}

\usepackage{amsmath}

\begin{document}

\title{Hamiltonian form of Carrollian gravity}

\author{Sandipan Sengupta}
\email{sandipan@phy.iitkgp.ac.in}
\affiliation{Department of Physics, Indian Institute of Technology Kharagpur, Kharagpur-721302, INDIA}

\begin{abstract}

We develop a Hamiltonian description of the `Carroll' (Levy Leblond-Sen Gupta) limit of gravity theory in the first-order formalism. Through a constraint analysis, the number of local degrees of freedom are shown to be two in this singular limit. The associated Hamiltonian constraint depends only on the densitized triad fields and their space derivatives. We also provide a scaling prescription within the Hilbert-Palatini theory to obtain the first-order analogue of the `electric' limit of canonical metric gravity. The simplicity of both these limiting Hamiltonian forms in these variables make them interesting candidates for quantization.

\end{abstract}

\maketitle

\section{Introduction}

The Lorentz transformations exhibit an intriguing singular limit opposite to the Galilean one. The corresponding set of boosts, discovered independently by Levy-Leblond \cite{ll} and Sen Gupta \cite{nd}, are given by:
\begin{eqnarray}
t'= t-\frac{x}{w}, ~x'=x,~y'=y,~z'=z,
\end{eqnarray}
where $w$ is a parameter with the dimension of velocity. These transformations are obtained by replacing the relative frame velocity by $\f{c^2}{w}$ in the original Lorentz transformations and then taking the limits of a small speed of light ($c<<w$) as well as small temporal separations between events ($c\Delta t<<\Delta x$). The parameter $w$, however, does not encode a physical velocity. Rather, it is the rate of motion $\f{\Delta x}{\Delta t}$ in the unprimed frame of an event that is fixed in time in the primed frame. These boosts, along with the spatial rotations and spacetime translations, form what has been christened the  `Carroll' group. As shown by Levy Leblond \cite{ll,bacry}, this ten-parameter group could be obtained through a In{\"o}n{\"u}-Wigner contraction of the Poincare group. 

Despite the nonstandard causal structure of the Levy Leblond-Sen Gupta spacetimes where the lightcone shrinks (almost) to a line, the Carrollian limit is found to be intimately connected to a remarkably wide range of physical contexts. For instance, a special case of this limit \cite{teit,henn} in the second order metric formulation of gravity  could be related to the BKL behaviour \cite{bkl} near a space-like singularity where time derivatives of fields are conjectured dominate over spatial ones. Carrollian physics is also relevant to the description of cosmological billiards \cite{nicolai}, null surfaces \cite{daut,leigh}, tachyons \cite{gibbons}, gravitational waves \cite{duval}, non-AdS holography \cite{flat}, inflationary cosmology \cite{infl}, multi-metric gravity \cite{zorba} and so on. 

The Levy Leblond-Sen Gupta spacetimes have been typically shown to arise in gravity theories where the metric is either exactly degenerate \cite{kaul} or becomes so in a limiting sense \cite{teit,henn}. The latter case has also been interpreted as a $c\rightarrow 0$ (or `strong coupling') limit. However, since $c$ has dimensions, such an interpretation demands caution as operationally it may lead to more than one inequivalent limits of the seed field theory, depending on the units chosen to define the fundamental variables. In fact, examples of such inequivalent limiting formulations of gravity have appeared in the earlier literature within approaches based on the gauging of Carroll algebra \cite{hartong,bergshoeff,guer} or through series expansions in powers of $c^2$ within a Lagrangian framework \cite{obers}. 

Recently, it has been claimed  that relativistic field theories in general could admit two inequivalent Carroll limits \cite{henneaux1}. This feature is exactly analogous to the `electric' and `magnetic' limits in Galilean electromagnetism \cite{ll1}, the basic variables in the two cases being assigned different units through relative factors of $c$.

Here, we develop a Hamiltonian formulation of Carrollian gravity within a first order framework, where the canonical variables as well as the constraint structure differ nontrivially from the metric (ADM) formalism. After solving the second-class constraints, we obtain a canonical theory corresponding to the so-called magnetic Carroll phase of metric gravity. We also explicitly demonstrate that the limiting theory exhibits the same number of physical degrees of freedom as in Hilbert-Palatini (or, Einstein-Hilbert) gravity, even though form of the limiting constraints are significantly different from the original ones.    

Next, by introducing a different scaling prescription motivated from the non-uniqueness of the limit $c\rightarrow 0$, we demonstrate how a first-order analogue of 
the `electric' limit of metric gravity could be formulated.

The final form of the Hamiltonian constraints in either limits are considerably simpler than the full constraint in pure gravity and are of interest from the canonical quantization perspective. 
Also, the latter case provides a remarkably simple Hamiltonian representation of the BKL conjecture \cite{bkl,sloan}. 

In the next section, we develop a canonical setup for the
magnetic limit following a brief outline of the Hamiltonian
structure of the Hilbert-Palatini theory. Based on an appro-
priate scaling limit, we obtain the Carroll phase of the
constraints without making any gauge choice. This is
followed by a constraint analysis based on their Poisson
algebra. The constraints in the time gauge are also found. In
Sec. III, we introduce a different scaling prescription and find
the resulting Hamiltonian theory. In time gauge, the asso-
ciated Hamiltonian formulation is found to be the first-order
counterpart of the electric limit in the metric formulation. The
concluding section contains a few relevant remarks.

\section{Limit A: Carroll gravity}
The first order or $SO(3,1)$ gauge formulation of Einstein gravity is based on the tetrad $e_\mu^I$ and spin-connection fields $\omega_\mu^{~IJ}$. The action principle involves the following Lagrangian density:
\begin{equation} \label{L0}
{\cal L}(\hat{e},\hat{\omega}) ~ = ~ \frac{1}{2\hat{\kappa}}\hat{e} \hat{e}^{\mu}_I \hat{e}^{\nu}_{J}\hat{R}_{\mu\nu}^{~~~ IJ}(\hat{\omega}) 
\end{equation}
where $\hat{\kappa}$ is the gravitational coupling, $\hat{e}\equiv \det{\hat{e}_\mu^I}$, $\hat{e}^{\mu}_I$ is the inverse tetrad field and  $\hat{R}_{\mu\nu}^{~~~ IJ}(\hat{\omega})=\del_{[\mu}\hat{\omega}_{\nu]}^{~IJ}+\hat{\omega}_{[\mu}^{~IK}\hat{\omega}_{\nu]K}^{~~~J}$ is the field strength. To begin with, let us consider the following limit \cite{bergshoeff} leading to a realization of the Carroll algebra, where the basic variables go to the new ones as:
\begin{eqnarray}\label{scaling1}
\hat{e}_\mu^0&=&\epsilon e_\mu^0,~\hat{e}_\mu^i= e_\mu^i,~\hat{\omega}_\mu^{~0i}=\epsilon \omega_\mu^{0i},~\hat{\omega}_\mu^{~ij}= \omega_\mu^{ij};\nonumber\\
\hat{e}^\mu_0&=&\f{1}{\epsilon} e^\mu_0,~\hat{e}^\mu_i= e^\mu_i.
\end{eqnarray}
The unhatted variables are assumed to remain finite in the limit $\epsilon\rightarrow 0$, implying that the determinant $\hat{e}$ approaches zero. In these redefined variables, the Hilbert-Palatini Lagrangian density becomes:
\begin{equation*} \label{LC}
{\cal L} ~ = ~ \frac{\epsilon}{2\hat{\kappa}}e e^{\mu}_I e^{\nu}_{J}R_{\mu\nu}^{~~~ IJ}(\omega) 
\end{equation*}
where the new field strengths $R_{\mu\nu}^{~~0i}=\bar{D}_{[\mu}\omega_{\nu]}^{~0i}:=\del_{[\mu}\omega_{\nu]}^{~0i}+\omega_{[\mu}^{~ik}\omega_{\nu]}^{~0k}$ and $\bar{R}_{\mu\nu}^{~~ ij}=\del_{[\mu}\omega_{\nu]}^{~ij}+\omega_{[\mu}^{~ik}\omega_{\nu]k}^{~~~j}$ represent the $SO(3)$ counterparts of the original ones ($\bar{D}_\mu$ denotes the SO(3) covariant derivative). Evidently, the Lagrangian density is well-defined provided the coupling scales as $\hat{\kappa}=\epsilon\kappa$. In what follows next, we set up a Hamiltonian formulation of the above limit  within the first-order framework.

We employ the standard Hamiltonian decomposition of the tetrad fields as \cite{peldan, kaul1}:
\begin{eqnarray*}
&& \hat{e}^{I}_{t} =  \sqrt{\hat{e}\hat{N}}\hat{M}^{I}+\hat{N}^{a}\hat{V}_{a}^{I} , ~ \hat{e}^{I}_{a} = 
\hat{V}^{I}_{a};\nonumber\\
&& \hat{M}_{I}\hat{V}_{a}^{I} = 0 , ~ \hat{M}_{I}\hat{M}^{I} =  -1;\nonumber\\
&&\hat{e}^{t}_{I}  =  -\frac{\hat{M}_{I}}{\sqrt{\hat{e}\hat{N}}} , ~ \hat{e}^{a}_{I} ~ = ~
\hat{V}^{a}_{I}+\frac{\hat{N}^{a}\hat{M}_{I}}{\sqrt{\hat{e}\hat{N}}}  ~ ~; \nonumber \\
&& \hat{M}^{I}\hat{V}_{I}^{a} :=  0 ,~ \hat{V}_a^I \hat{V}^b_I ~ := ~ \delta_a^b ,~ \hat{V}_a^I
\hat{V}^a_J:= \delta^I_J + \hat{M}^I \hat{M}_J .
\end{eqnarray*} 
The spatial metric and its determinant are defined as: $ \hat{q}_{ab} ~ := ~\hat{V}_{a}^{I}\hat{V}_{bI}$ and $\hat{q} := \mathrm{det}\hat{q}_{ab}\
$, with $\hat{e} := det(\hat{e}^{I}_{\mu}) = \hat{N}\hat{q}$. Using the identity,
\begin{equation*}
\hat{e}_{I}^{[a}\hat{e}_{J}^{b]} ~ = ~  \hat{N} \hat{e} \hat{e}^{t}_{[I} \hat{e}^{[a}_{K]} \hat{e}^{b]}_{[J} \hat{e}^{t}_{L]}\eta^{KL}
+ \hat{N}^{[a}\hat{e}^{b]}_{[I} \hat{e}^{t}_{J]} ,
\end{equation*}
the original Lagrangian density (\ref{L0}) becomes:
\begin{eqnarray}
{\cal L}&=&\f{\hat{e}}{\kappa}[\hat{e}^t_I \hat{e}^a_J \hat{R}_{ta}^{~~IJ}+\f{1}{2}( \hat{N} \hat{e} \hat{e}^{t}_{[I} \hat{e}^{[a}_{K]} \hat{e}^{b]}_{[J} \hat{e}^{t}_{L]}\eta^{KL}\nonumber\\
&&+ \hat{N}^{[a}\hat{e}^{b]}_{[I} \hat{e}^{t}_{J]})\hat{R}_{ab}^{~~IJ}]
\end{eqnarray}
Defining the momenta conjugate to $\hat{\omega}_a^{~IJ}$ as $\hat{\Pi}^a_{~IJ}:=\f{\hat{e}}{\hat{\kappa}}\hat{e}^t_{[I}\hat{e}^a_{J]}$, we may rewrite the above as:
\begin{eqnarray}
{\cal L}=\f{1}{2}\hat{\Pi}^a_{~IJ}\del_t \hat{\omega}_a^{~IJ}-{\cal H}
\end{eqnarray}
where the Hamiltonian density ${\cal H}$ is completely constrained owing to the absence of velocities corresponding for the fields $\hat{N},~\hat{N^a},~\hat{\omega}_t^{~IJ}$:
\begin{eqnarray}
{\cal H}=\hat{N}\hat{H}+\hat{N}^a\hat{H}_a+\f{1}{2}\hat{\omega}_t^{~IJ}\hat{G}_{IJ}
\end{eqnarray}
The constraints read:
\begin{eqnarray}\label{fullc}
\hat{H}&=&\f{\hat{\kappa}}{2}\hat{\Pi}^a_{~IK}\hat{\Pi}^b_{~JL}\eta^{KL} \hat{R}_{ab}^{~~IJ}~\approx 0,\nonumber\\
 ~\hat{H}_a&=&\f{1}{2}\hat{\Pi}^b_{~IJ} \hat{R}_{ab}^{~~IJ}~\approx 0,\nonumber\\
 \hat{G}_{IJ}&=&-\hat{D}_a \hat{\Pi}^a_{~IJ}~\approx 0
\end{eqnarray}
In addition, the momenta are associated with six primary constraints, reflecting that only twelve of its components are independent:
\begin{eqnarray}
\hat{C}^{ab}:=\f{1}{2}\epsilon^{IJKL}\hat{\Pi}^a_{~IJ}\hat{\Pi}^b_{~KL}\approx 0
\end{eqnarray}

We may implement the Carroll limit (\ref{scaling1}) through the following redefinitions of the canonical variables, where, as earlier, the unhatted variables remain finite in the limit $\epsilon \rightarrow 0$:
\begin{eqnarray}\label{scaling2}
\hat{\omega}_\mu^{~0i}&=&\epsilon  \omega_\mu^{~0i},~\hat{\omega}_\mu^{~ij}= \omega_\mu^{~ij},~\hat{\Pi}^a_{~0i}=\f{1}{\epsilon} \Pi^a_{~0i},~\hat{\Pi}^a_{~ij}= \Pi^a_{~ij};\nonumber\\
\hat{N}&=&\epsilon N,~ \hat{N}^a=N^a.
\end{eqnarray}
Note that these scalings preserve the symplectic form, as is necessary. The transformation of the lapse follows from the scaling of the tetrad determinant. Let us now analyse the Hamiltonian structure of gravity theory resulting from the above Carroll limit.

The full Hamiltonian constraint in the new variables reads:
\begin{eqnarray*}
\hat{H}=\f{\epsilon \kappa}{2}\left[-\f{1}{\epsilon^2}\Pi^a_{~0i} \Pi^b_{~0j} R_{ab}^{~~ij}+\Pi^a_{~ik} \Pi^b_{~jk} R_{ab}^{~~ij}+2\Pi^a_{~0k} \Pi^b_{~ik} R_{ab}^{~~0i}\right]
\end{eqnarray*}
where the new field strength decomposes into its $SO(3)$ counterpart $\bar{R}_{ab}^{~~ij}$ and the rest: $R_{ab}^{~~ij}=\bar{R}_{ab}^{~~ij}+\epsilon^2 \omega_{[a}^{~0i}\omega_{b]}^{~0j}$. Redefining the lapse as in (\ref{scaling2}), then using $\hat{N}\hat{H}=NH$ and finally implementing the limit $\epsilon\rightarrow 0$, we obtain from the above:
\begin{eqnarray}
H=-\f{\kappa}{2}\Pi^a_{~0i} \Pi^b_{~0j} \bar{R}_{ab}^{~~ij}
\end{eqnarray}
The right hand side is clearly well-defined since the variables are all finite in the limit. An immediate fact to be gleaned from the above is the vanishing Poisson bracket $[H,H]=0$, a characteristic of the Carroll algebra.

The new vector constraint also retains only the $SO(3)$ covariant derivative and field-strength in the limit:
\begin{eqnarray}
H_a
=\Pi^b_{~0i} \bar{D}_{[a}\omega_{b]}^{~0i}+\f{1}{2}\Pi^b_{~ij} \bar{R}_{ab}^{~~ij}
\end{eqnarray}
Proceeding similarly, we find the following limiting forms of the rotation and boost constraints:
\begin{eqnarray}
G_{ij}=-\bar{D}_a \Pi^a_{~ij}+\omega_{a0[i} \Pi^{a0}_{~~j]},~
G_{0i}=-\bar{D}_a \Pi^a_{~0i}
\end{eqnarray}
Evidently, the rotation constraint is the only one that exhibits no change when compared with the full Hilbert-Palatini constraints.

\subsection{Algebra of Carrollian constraints}
Following a tedious but straightforward analysis, we summarize the algebra of the Carrollian constraints below:
\begin{eqnarray}\label{poisson}
&&\left[\int MH,\int N^a H_a\right]=\int [\left(M\del_a N^a-N^a\del_a M\right)H\nonumber\\
&&~~+ M N^a \Pi^b_{~0j}\bar{R}_{ab}^{~~ij}G^{0i}],\nonumber\\
&&\left[\int N^a H_a,\int M^b H_b\right]=\int [\left(M^d\del_d N^b-N^d\del_d M^b\right)H_b\nonumber\\
&&~~ -N^a M^b R_{ab}^{~~IJ} G_{IJ}],\nonumber\\
&&\left[\int MH,\int \Lambda_{IJ}G^{IJ}\right]= 0=\left[\int M^a H_a,\int \Lambda_{IJ}G^{IJ}\right],\nonumber\\
&&\left[\int \Lambda_{IJ}G^{IJ},\int \Lambda_{KL}G^{KL}\right]=4\int \Lambda_{I}^{~~K} \Lambda_{KJ}G^{IJ},\nonumber\\
&&\left[\int M^a H_a,\int \lambda_{cd}C^{cd}\right]=\int [(M^c\del_c \lambda_{cd}+2\lambda_{ac}\del_a M^c\nonumber\\
&&~~-\lambda_{ab}\del_c M^c)C^{ab}-\f{1}{2}\epsilon_{ijk}\lambda_{ab}N^a\left(\Pi^{b0i}G^{jk}+\Pi^{cjk}G^{0i}\right)]\nonumber\\
&&\left[\int MH,\int \lambda_{ab}C^{ab}\right]=2\int M\lambda_{ab}\epsilon^{ijk}[\Pi^c_{~0k}\Pi^b_{~0j}\bar{D}_c\Pi^a_{~0i}\nonumber\\
&&~~+\Pi^b_{~0k}\Pi^a_{~0i} G_{0j}]
\end{eqnarray}
The last equality leads to a secondary constraint:
\begin{eqnarray}
D^{ab}:=\epsilon^{ijk}\Pi^c_{~0k}\Pi^{(b}_{~0j}\bar{D}_c\Pi^{a)}_{~0i}\approx 0,
\end{eqnarray}
beyond which there is no further one. With this, the only second-class pair is given by $(C^{ab},D^{cd})$:
\begin{eqnarray*}
\left[\int \lambda_{ab}C^{ab},\int \mu_{cd}D^{cd}\right]=4 \int \lambda_{a[b}\mu_{d]c} \Pi^a_{~0i}\Pi^b_{~0k}\Pi^c_{~0k}\Pi^d_{~0i}.
\end{eqnarray*}

\subsection{Solution of second-class constraints}
The second-class constraints could be solved by eliminating the six redundant variables in $\hat{\Pi}^a_{~IJ}$ through the parametrization $\Pi^a_{~0i}=E^a_i,~\Pi^a_{~ij}=\chi_{[i}E^a_{j]}$, following a standard approach in first order gravity \cite{peldan}. The momenta $E^a_i$ essentially turn out to be the densitized triad, as could be seen from the definition of $\Pi^a_{~IJ}$. The resulting symplectic form now depends only on the twelve independent  canonical pairs:
\begin{eqnarray}
\Omega=\f{1}{2}\Pi^a_{~IJ}\del_t\omega_a^{~IJ}=E^a_i\del_t Q_a^i+\zeta_j \del_t \chi^j
\end{eqnarray}
where we have introduced the new fields as: $Q_a^i=\omega_a^{~0i}-\chi_j \omega_a^{~ij},~\zeta^j=E^a_i \omega_a^{~ij}$. The last equality also implies that only three components ($\zeta^i$) of $\omega_a^{~ij}$
are dynamical, allowing for the following parametrization:
\begin{eqnarray}
\omega_a^{~ij}=\frac{1}{2}E_a^{[i}\zeta^{j]}+\epsilon^{ijk} E_a^l N^{kl}
\end{eqnarray}
Clearly, the velocities of the six fields $N_{kl}=N_{lk}$ do not appear in the Lagrangian density. The vanishing of their conjugate momenta ($P^{kl}\approx 0$) leads to a further set of secondary constraints:
\begin{eqnarray}
[H,P^{kl}]\approx 0 
\end{eqnarray}
Solution to these constraints (along with the boost) is equivalent to the vanishing of (spatial) torsion. This also implies that the fields $\omega_a^{~ij}$ are completely determined in terms of the momenta $E^a_i$.

Following the solution of the second-class constraints as above, the Carrollian family of first-class constraints in the new variables finally become:
\begin{eqnarray}\label{const}
H&=&-\f{\kappa}{2} E^a_i E^b_j \bar{R}_{ab}^{~~ij}(E)~\approx 0,\nonumber\\
H_a&=&E^b_i \left[\bar{D}_{[a}Q_{b]}^i+\omega_{[b}^{~ik}\del_{a]}\chi_k\right]~\approx 0,\nonumber\\
G_{ij}&=&-\left[\del_a(\chi_{[i} E^a_{j]})+\chi_{[j} \zeta_{i]}-Q_{a[i}E^a_{j]}\right]~\approx 0\nonumber\\
G_{0k}&=& -\del_a E^{a}_k+\zeta_k~\approx 0.
\end{eqnarray}

\subsection{Time gauge}
We may simplify the theory even further by setting the time gauge $\chi_i=0$, a condition which forms a second-class pair with the boost constraint given by eq.(\ref{const}). In this gauge, the solution of boost thus turns the associated conjugate momenta into a dependent variable as: $\zeta_i=\del_a E^a_i$. 

Let us emphasize that in order to obtain the Carrollian constraints directly in time gauge, the limit should be taken after implementing the gauge. However, in this particular case here, this procedure is equivalent to taking the limit before imposing the gauge.

In this gauge, the Carroll limit of the constraints could be read off from eq.(\ref{fullc}) as:
\begin{eqnarray}
H &=&-\f{\kappa}{2} E^a_i E^b_j \bar{R}_{ab}^{~~ij}(E)~\approx 0,\nonumber\\
H_a &=& E^b_i \bar{D}_{[a} Q_{b]}^i~\approx 0,\nonumber\\
G_{ij}&=& Q_{a[i} E^a_{j]}~\approx 0
\end{eqnarray}
These form a system of first-class constraints, while obeying the Carroll algebra.

With these set of constraints, one may directly proceed towards a canonical quantization. 

Let us add that the rotation constraint in time gauge could be solved as $Q_a^k=E_a^l M^{kl}$ where $M_{kl}=M_{lk}$ is an arbitrary field. This indeterminacy of the connection $\omega_a^{~0i}$ resembles a similar feature found in the solutions of gravity theory where the tetrad fields are degenerate in an exact sense \cite{kaul} (as opposed to the limiting sense as here). There it had been demonstrated that the Lagrangian equations of motion leave both $\omega_a^{~0i}$ and $\omega_a^{~ij}$ undetermined upto two symmetric fields $M^{kl}$ and $N^{kl}$. On the other hand, this also shows that the degenerate limit of gravity is not really equivalent to a gravity theory based on exactly degenerate tetrad fields. 

We may also note that this ambiguity upto $M_{kl}$ is equivalent to the one appearing within a Lagrangian description of the `magnetic' Carroll limit of gravity \cite{bergshoeff}, where a similar field shows up as a multiplier in the second order action. 

\subsection{Physical degrees of freedom}
Based on the Poisson algebra presented earlier, it is straightforward to count the physical degrees of freedom in the limiting Hamiltonian theory. Before setting the time gauge, the canonical pairs ($\omega_a^{~IJ},\Pi^a_{~IJ}$) are subject to 10 first class constraints $H,~H_a,~G_{0k},~G_{ij}$ and six second-class pairs ($C^{ab},~D^{cd}$). This leaves two degrees of freedom per spacetime point in the Carroll phase of gravity. This is same as in Hilbert-Palatini gravity, despite the fact that the first case represents a singular limit of the latter and the explicit canonical form and details of the constraints exhibit nontrivial differences between the two cases. A counting after fixing the time gauge leads to the same result.

\section{Non-uniqueness of the $c\rightarrow 0$ limit}
It has already been emphasized earlier that a formal limit defined through the dimensionful parameter $c$ need not be unique. In this section we show that it is possible to invoke a different scaling prescription for the first order variables based on the limit $c\rightarrow 0$. As it turns out, this results in a first-order analogue of the so-called `electric Carroll limit' of gravity in time gauge.

\subsection{A different scaling: Limit B} 
Let us consider the identity: $-1=\frac{1}{c^2} \hat{e}_{\mu 0} \hat{e}^{\mu}_0=\left[\frac{1}{c}\hat{e}_{\mu}^0\right]\left[c\hat{e}^{\mu}_0\right]$, where the appropriate factors of $c$ have been restored in the internal metric $\eta_{IJ}$. A possible way to define a singular limit from the above is to keep the factors $\f{1}{c}\hat{e}_\mu^0:=e_\mu^{0}$ and $c\hat{e}^\mu_0=e^{\mu}_0$ finite as $c\rightarrow 0$, which implies:
\begin{eqnarray}
\hat{e}_\mu^0\rightarrow 0,~\hat{e}^\mu_0 \rightarrow \infty.
\end{eqnarray}
This is precisely the scaling prescription (\ref{scaling1}) discussed at the earlier section.

The other alternative, however, is to change the units of the basic variables and keep the factors $c\hat{e}_\mu^0:=e_\mu^{0}$ and $\f{1}{c}\hat{e}^\mu_0=e^{\mu}_0$ finite instead, as $c\rightarrow 0$. This is what we consider here. 

To understand how the two sets of variables subject to the two different scalings above are mutually related, let us define a new set of singular variables as: 
\begin{eqnarray}
\hat{e}_\mu^{0}=c^2 \tilde{e}_\mu^{0},~\hat{e}_\mu^{i}= \tilde{e}_\mu^{i},~\hat{\omega}_\mu^{~0i}=c^2\tilde{\omega}_\mu^{~0i},~\hat{\omega}_\mu^{~ij}=\tilde{\omega}_\mu^{~ij}
\end{eqnarray}
Next, using the relations between the singular (hatted) and nonsingular (unhatted) variables given in (\ref{scaling1}), and rewriting the $c\rightarrow 0$ limit through the dimensionless parameter $\epsilon$ as $c=\tilde{c}\epsilon$ ($\tilde{c}$ being a finite characteristic speed), we obtain:
 \begin{eqnarray}\label{scalingB1}
&&\tilde{e}_\mu^0=\f{1}{\epsilon} e_\mu^0,~\tilde{e}_\mu^i= e_\mu^i,~\tilde{e}^\mu_0=\epsilon e^\mu_0,~\tilde{e}^\mu_i= e^\mu_i,\nonumber\\
&&\tilde{\omega}_\mu^{~0i}=\frac{1}{\epsilon}\omega_\mu^{~0i},~\tilde{\omega}_\mu^{~ij}=\omega_\mu^{~ij},
\end{eqnarray} 
where we have redefined the finite (in the limit $\epsilon\rightarrow 0$) variables by absorbing factors of $\tilde{c}^2$. The above imply that the canonical momenta now scale as: $\tilde{\Pi}^a_{0i}=\epsilon \Pi^a_{0i},~\tilde{\Pi}^a_{ij}= \Pi^a_{ij}$, and the determinant of the inverse tetrad ($\det\tilde{e}^\mu_I$) vanishes in the limit.

Next, we obtain all the constraints in the Hamiltonian theory starting from  the Hilbert-Palatini Lagrangian density in the new variables and set the time gauge. In order to avoid repetitions, we display the new constraints directly in the nonsingular variables using the same procedure elucidated in the earlier section. 

%
%


\subsection{Constraints}
In time gauge the boost constraint is already solved. Along with the vanishing torsion condition, this implies that the spatial connection $\omega_a^{~ij}(E)$
is to be treated as a functional of the momenta $E^a_i$.
We thus obtain:
\begin{eqnarray}
H&=&-\f{\kappa}{2}E^a_{i} E^b_{j} \left[\epsilon^2 \bar{R}_{ab}^{~~ij}(E)+Q_{[a}^{i}Q_{b]}^{j}\right]~\approx 0,\nonumber\\
H_a&=&E^b_{i} \bar{D}_{[a}Q_{b]}^{i}~\approx 0,\nonumber\\
G_{ij}&=&Q_{a[i} E^{a}_{j]}~\approx 0
\end{eqnarray} 
Now we can impose the limit $\epsilon\rightarrow 0$, which changes the Hamiltonian constraint to the one below while  preserving the rest:
\begin{eqnarray}\label{H}
H&=&-\f{\kappa}{2}E^a_{i} E^b_{j} Q_{[a}^{i}Q_{b]}^{j}~\approx 0
\end{eqnarray}
With this, we obtain a Hamiltonian theory inequivalent to the one in limit A in the earlier section. This readily reflects the fundamental feature of Carroll algebra: $[H,H]=0$. 

The Hamiltonian constraint (\ref{H}) could be identified with the one in second order metric gravity (corresponding to different canonical variables and constraint structure), where only the extrinsic curvature dependent terms survive and the spatial curvature term drops out, called a `strong-coupling' or `zero-signature' or `electric' limit in the literature \cite{teit,henn,hartong}.

\section{Conclusions}
Here, we have set up a Hamiltonian formulation of the singular `Carroll' limit of Hilbert-Palatini gravity. Through a detailed constraint analysis, we find that the number of physical degrees of freedom (locally) is two, which is the same as in the original theory. This answers an important and oft quoted open question \cite{hartong,bergshoeff} in the literature regarding the number of polarization states in `Carroll' gravity.

The Carrollian Hamiltonian constraint in limit A (so called `magnetic' limit)  is free from any ordering ambiguity. It depends only on the momenta ($E^a_i$) and their spatial derivatives. Owing to this simplication compared to the original Hilbert-Palatini gravity, this is an interesting candidate Hamiltonian to quantize.

Furthermore, we provide a scaling prescription (limit B) based on the Levy-Leblond-Sen Gupta limit which naturally leads to a canonical form inequivalent to the first one (limit A). The associated constraints represent a first-order analogue of the so-called `electric' limit of metric gravity. Limit B is expected to be relevant in the study of the celebrated BKL conjecture from a Hamiltonian viewpoint. Even though the corresponding Hamiltonian constraint, when quantized, exhibits an ordering ambiguity, it is purely algebraic (contains no derivatives of coordinate or momenta). Again, this is quite an attractive feature, as compared to the full Hamiltonian constraint of Hilbert-Palatini gravity. A quantization of this Hamiltonian theory could lead to new insights regarding a possible connection between quantum gravity and a potential resolution of gravitational singularities \cite{rc,hawking}. 

To sum up, the Hamiltonian theories in the limits A and B both provide an intriguing arena for quantum gravity, modulo issues such as operator ordering, regularization or interpretational aspects.

As a final remark, our analysis also shows that the Carrollian limit of gravity is in general inequivalent to a gravity theory based on an exactly degenerate tetrad. In fact, gravity theory in vacuum based on a noninvertible tetrad (metric) has recently been shown to be associated with three local degrees of freedom \cite{seng}, reflecting a discontinuity in the limit to a vanishing tetrad determinant. Let us also observe that only one of the two undetermined symmetric fields which had been found to appear in the generic spacetime solutions with noninvertible tetrads \cite{kaul} shows up in the Carrollian phase. It is plausible that there exist further limiting phases of Hilbert-Palatini gravity connected to the ones considered here, where the other symmetric field found in ref.\cite{kaul} could enter the description. Some of these
intriguing questions are worth a deeper study. \footnote{After the completion of this work, we noted the eprint in \cite{henn2}, aiming to demonstrate the equivalence of the first and second order actions associated with the `magnetic' limit. 
While it is not concerned with a Hamiltonian analysis involving the first order constraints (without fixing any gauge) as here, there appears a brief discussion on the `magnetic' limit in time gauge whose results have a partial overlap with ours here in one of the subsections. Our perspective and approach, however, are different.}.

\begin{acknowledgments}
Support (in part) of the SERB, DST, Govt. of India., through the MATRICS project grant MTR/2021/000008 is gratefully acknowledged. I also thank Marc Henneaux for helpful clarifications on ref.\cite{henneaux1} through a private communication.
\end{acknowledgments}


\begin{thebibliography}{99}

\bibitem{ll} J. M. Levy-Leblond, Ann. Inst. Henri Poincare, 3 (1965), 1.

\bibitem{nd} N. D. Sen Gupta, Nuovo Cimento 44 (1966), 512.

\bibitem{bacry} H. Bacry, J. M. Levy-Leblond, J. Math. Phys. 9 (1968), 1605-1614.

\bibitem{teit} C. Teitelboim, “Surface deformations, their square root and the signature of space-time”
Print-78-1134 (Princeton), published in: Austin Group Theor.(1978): 362; Contribution to:
7th International Group Theory Colloquium: The Integrative Conference on Group Theory
and Mathematical Physics;\\
C. Teitelboim, Annals Phys. 79 (1973), 542-557.

\bibitem{henn} M. Henneaux, Bull. Soc. Math. Belg. 31 (1979),
47-63 Print-79-0606 (Princeton).

\bibitem{bkl} V. A. Belinsky, I. M. Khalatnikov and E. M. Lifshitz, Adv. Phys. 19 (1970), 525-573;\\
V. A. Belinsky, I. M. Khalatnikov and E. M. Lifshitz, Adv. Phys. 31 (1982), 639-667.

\bibitem{nicolai} T. Damour, M. Henneaux, H. Nicolai, Class. Quant. Grav. 20 (2003), R145-R200.

\bibitem{daut} G. Dautcourt, Acta Phys. Polon. B 29 (1998), 1047-1055.

\bibitem{leigh} L. Ciambelli, R. G. Leigh, C. Marteau and P. M. Petropoulos, Phys. Rev. D 100 (2019), 046010;\\
 L. Donnay and C. Marteau, Class. Quant. Grav. 36 (2019), 165002.
 
\bibitem{gibbons} G. Gibbons, K. Hashimoto, P. Yi, JHEP 09 (2002), 061.

 
 \bibitem{duval} C. Duval, G. W. Gibbons, P. A. Horvathy, P.-M. Zhang, Class. Quant. Grav. 34 (2017), 175003.

\bibitem{flat} L. Ciambelli, C. Marteau, A. C. Petkou, P. M. Petropoulos and K. Siampos, JHEP 07 (2018), 165;\\
A. Bagchi, R. Basu, A. Kakkar and A. Mehra, JHEP 12 (2016), 147.


\bibitem{infl} J. de Boer, J. Hartong, N. A. Obers, W. Sybesma and S. Vandoren, Front. in Phys. 10 (2022), 810405.

\bibitem{zorba} E. Ekiz et. al., JHEP 10 (2022) 151.

\bibitem{kaul} R. K. Kaul and S. Sengupta, Phys. Rev. D 93 (2016), 084026;\\
R. K. Kaul and S. Sengupta, Phys. Rev. D 94 (2016), 104047.



\bibitem{hartong} J. Hartong, JHEP 08 (2015), 069 .

\bibitem{bergshoeff} E. Bergshoeff, J. Gomis, B. Rollier, J. Rosseel and T. ter Veldhuis, JHEP 03 (2017), 165.

\bibitem{guer} A. Guerrieri, R. F. Sobreiro, Class. Quant. Grav. 38 (2021), 245003;\\
J. Figueroa-O'Farrill, E. Have, S. Prohazka, J. Salzer, arXiv:2206.14178 [hep-th].

\bibitem{obers} 
D. Hansen, N. A. Obers, G. Oling and B. T. Søgaard, arXiv:2112.12684 [hep-th].

\bibitem{henneaux1} M. Henneaux and P. Salgado-Rebolledo, JHEP 11 (2021), 180.

\bibitem{ll1} M. Le Bellac and J.-M. Levy-Leblond, Nuovo Cimento B 14 (1973), 217.

\bibitem{sloan} A. Ashtekar, A. Henderson, D. Sloan, Class. Quant. Grav. 26 (2009) 052001;\\
A. Ashtekar, A. Henderson, D. Sloan, Phys. Rev. D 83 (2011) 084024.



\bibitem{peldan} P. Peldan, Class. Quant. Grav. 11 (1994), 1087-1132;\\
 N. Barros e Sa, Int. J. Mod. Phys. D10 (2001), 261-272.

\bibitem{kaul1}  G. Date, R. K. Kaul, S. Sengupta, Phys. Rev. D 79 (2009), 044008;\\
 R. K. Kaul, S. Sengupta, Phys. Rev. D 85 (2012), 024026.

\bibitem{rc} A. Raychaudhuri, Phys. Rev. 98 (1955), 1123.

\bibitem{hawking} S. W. Hawking, Proc. R. Soc. Lond. A 294 (1966), 511.

\bibitem{seng} S. Sengupta, arXiv:2210.15980 [gr-qc].

\bibitem{henn2} A. Campoleoni, M. Henneaux, S. Pekar, A. Perez, P. Salgado-Rebolledo, arxiv:2207.14167 [hep-th].

\end{thebibliography}
\end{document}